\documentclass{ws-p8-50x6-00}
\begin{document}

\title{Ultra-Peripheral Collisions in STAR}
\author{P. Yepes}
\address{ for the STAR Collaboration \\Rice University\\ 
MS 315, 6100 Main Street, Houston, TX 77005 \\ E-mail: yepes@rice.edu}

\maketitle

\abstracts{ Ultra-peripheral heavy ion  collisions involve long range electromagnetic
interactions at impact parameters larger than twice the nuclear radius,
where no hadronic nucleon-nucleon collisions occur. 
The first observation of coherent $\rho^0$ production  with and without accompanying nuclear 
breakup, along with the observation of $e^+e^-$ pair production  
are reported by the STAR collaboration.}

\section{Introduction}
\begin{figure}[!b]
\vspace*{-.7cm}
\begin{center}
\leavevmode
\epsfclipon
\epsfysize=3.5  cm
\epsfbox{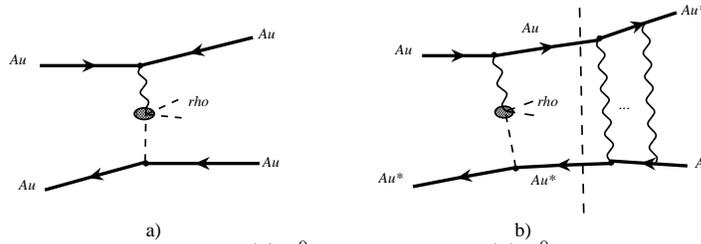}
\end{center}
\vskip -.25 in
\caption[]{Feynman diagrams for (a) $\rho^0$ production, and (b)
$\rho^0$ production with nuclear excitation.
Factorization is indicated by the dashed line; nuclear excitation 
may occur by the exchange of a single or multiple photons. }
\label{fig:feynman}
\end{figure}
We refer to ultra-peripheral heavy ion collisions, as those nuclear 
interactions with a impact parameter, $b$, larger than twice the nuclear 
radius $R_A$, where no hadronic nucleon-nucleon collisions occur~\cite{baurrev}. 
The large charge of a relativistic heavy nucleus is a 
strong source of quasi-real photons. Exclusive $\rho^0$ meson production
$AuAu\! \rightarrow\! Au Au \rho^0$ can then  be described by the  vector meson 
dominance model~\cite{sakurai}:
a photon emitted by one nucleus fluctuates 
to a virtual quark-anti quark  pair; this intermediate state scatters 
diffractively 
from the other nucleus, emerging  as a vector meson. 
The diagram for this process is shown in Fig.~\ref{fig:feynman}a. 
Here, the gold nuclei remain  in their ground state.
Additional photon exchange can yield nuclear excitation and the subsequent
emission of single or multiple neutrons, as shown in Fig.~\ref{fig:feynman}b 
for the process  $AuAu \rightarrow Au^\star Au^\star \rho^0$.

Photon and Pomeron can couple coherently to the spatially extended
electric and nuclear charge of the gold nuclei. Requiring a coherent 
interaction of the whole nucleus imposes a maximum momentum
transfer of $\hbar / d$, where $d$ is the nuclear dimension. 
In the transverse direction along which nuclei are not Lorentz contracted,
$d$ is of the order of the nucleus radius, and therefore
$p_T \!<\! \hbar/ R_A$ ($\!\sim\!100$~MeV/c for $R_A \sim 7$fm).
In the longitudinal direction nuclei are contracted by a factor $\gamma$,
therefore the maximum longitudinal momentum is $\gamma$ times larger:
$p_\| \!< \! \hbar \gamma / R_A$ ($\!\sim\!6$~GeV/c at RHIC).

The coupling strength of the photon is proportional to  the square of the 
charge $Z^2$ (6241 for $AuAu$);  the strength of the Pomeron coupling lies 
between $A^{4/3}$ for surface coupling to $A^2$ in the  bulk limit 
($10^3$ to $10^4$ for $AuAu$). For gold collisions at
$\sqrt{s_{NN}}=130$ GeV those large couplings translate into a 
$\rho$ production cross section of about
400 mb, or $\approx$5\% of the total hadronic cross section.

In the case of $\rho$ production, it is not possible to determine 
which nucleus is the photon source and which is the target.
Therefore the amplitudes for $\rho^0$ production from both ions interfere.
Since the $\rho^0$ has negative parity, this interference is  destructive.
The short-lived $\rho^0$ decay before they travel the distance of the impact 
parameter $b$, and the interference is believed to be  sensitive to the 
post-decay wave function\cite{interfere}. 

\begin{figure}[!ht]
\resizebox{.45\textwidth}{!}
{\includegraphics[height=.5\textheight,angle=180,clip=true]{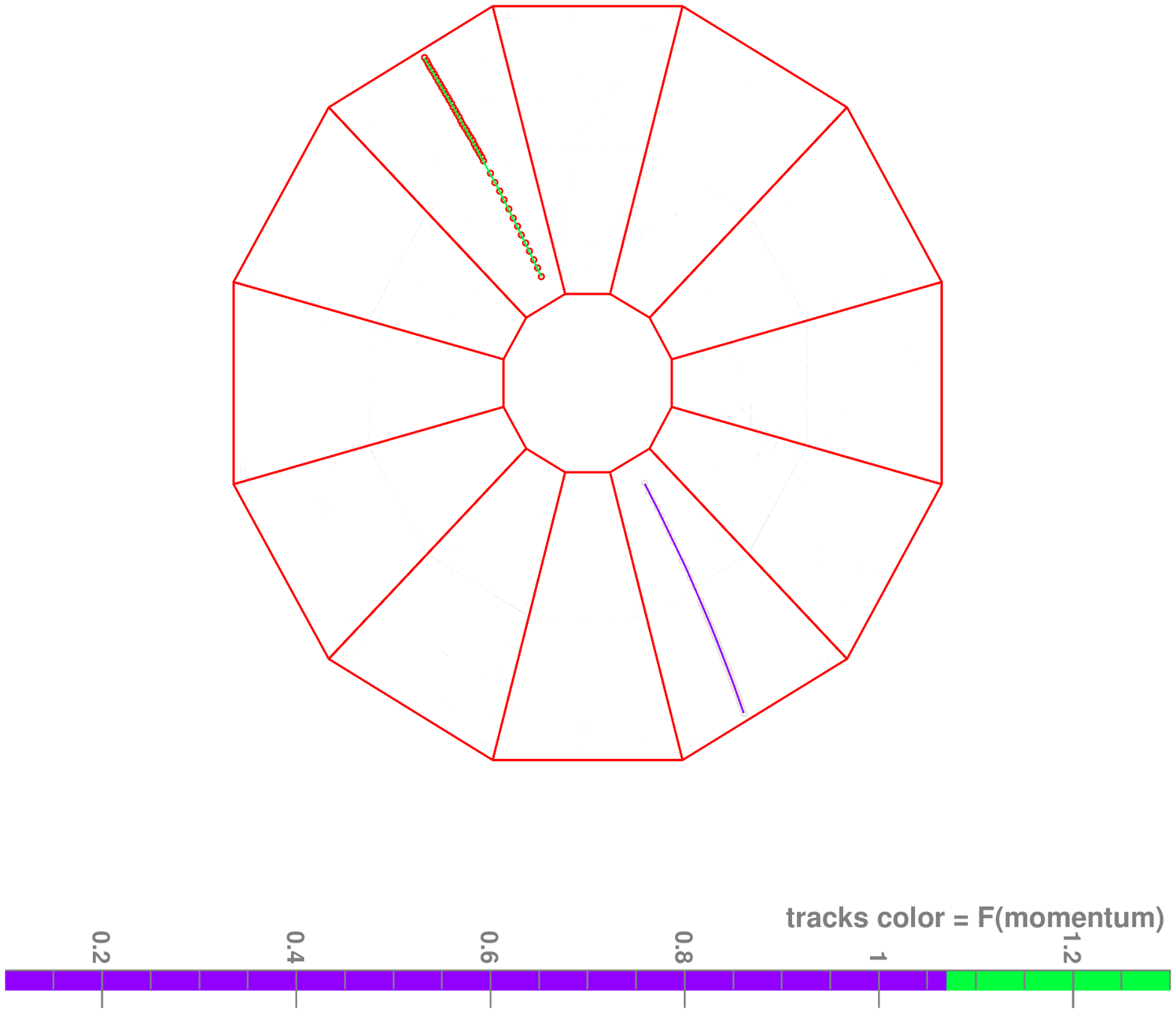}}
\resizebox{.45\textwidth}{!}
{\includegraphics[height=.5\textheight,angle=180,clip=true]{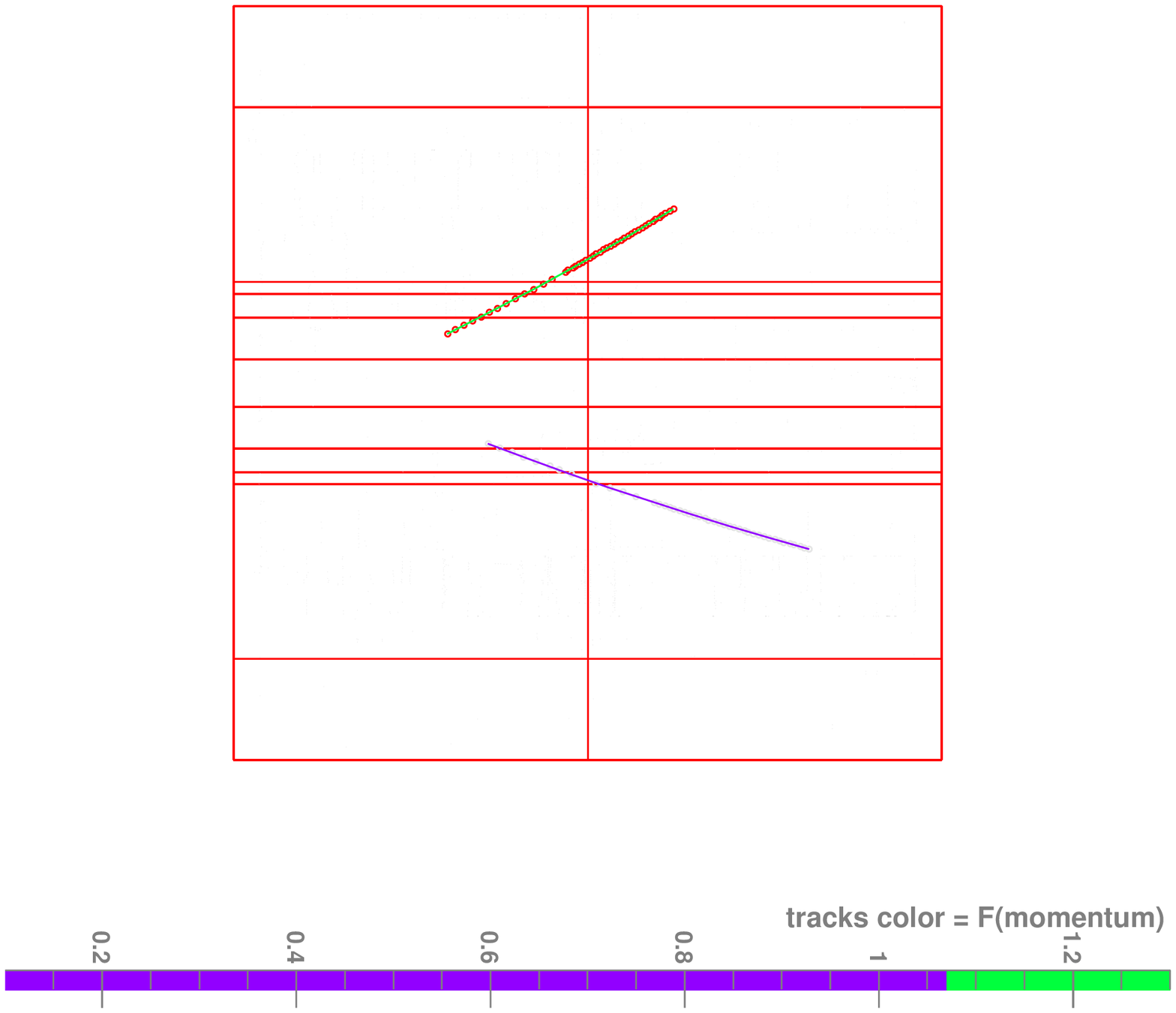}}
\label{fig:event}
\caption{End and side views of a typical $\rho$ candidate event in
the STAR TPC.  The candidate tracks are almost back-to-back radially,
but boosted longitudinally.}
\end{figure}

The $\rho$ meson decays into $\pi\pi$ with an approximately 100\% branching
ration. However a $\pi\pi$ final state may be produced directly or through the $\rho$.
The amplitudes for $\rho$ production, $A$, and direct $\pi\pi$
production, $B$, interfere and\cite{soding}
\begin{equation}
{d\sigma\over dM_{\pi\pi}} = \bigg| 
{A\sqrt{M_{\pi\pi}M_\rho \Gamma_\rho} \over M_{\pi\pi}^2 - M_\rho^2
+iM_\rho\Gamma_\rho} + B \bigg|^2
\label{eq:rhosigma}
\end{equation}
where the $\rho$ width is corrected for the increasing phase space as
$m_{\pi\pi}$ increases.  The $\rho$ component undergoes a 180$^o$
phase shift at $M_\rho$, so the interference skews the $\rho$ peak
shape, enhancing production for $M_{\pi\pi} < M_\rho$ and suppressing
the spectrum for $M_{\pi\pi} > M_\rho$.

\section{STAR Detector}
In the year 2000, RHIC collided 
gold nuclei at a center-of-mass energy of $\sqrt{s_{NN}}\!=\! 130$~GeV/nucleon. 
The STAR detector consists of a 4.2~m long cylindrical time projection chamber (TPC) 
of  2~m radius. In 2000 the TPC was operated
in a 0.25~T  solenoidal magnetic field.
Particles are  identified by their energy loss in the TPC.
A central trigger barrel of scintillators surrounds the TPC, providing
fast signaling for online event selection. 
Two zero degree calorimeters (ZDC) at $\pm$ 18m  from the interaction point 
are sensitive to the neutral  remnants of  nuclear breakup.

\section{Analysis}
Production of the $\rho^0$ meson in ultra-peripheral collisions has a clear
experimental signature, especially when compared with the typical high
multiplicity AuAu interaction. It is characterized by a $\pi^+\pi^-$ pair
in an otherwise 'empty' spectrometer. Fig.~\ref{fig:event} shows a  typical event 
candidate; the tracks are approximately  back-to-back in the transverse plane
due to the  small $p_T$ of the pair. 
Two data sets are used for the analysis: the 'minimum bias' and the
'topology' samples.

\begin{figure}[!h]
\begin{minipage}{0.45\textwidth}
\vspace*{-0.6cm}
\hspace*{-0.2\textwidth}
{\includegraphics[height=.4\textheight,clip=true]{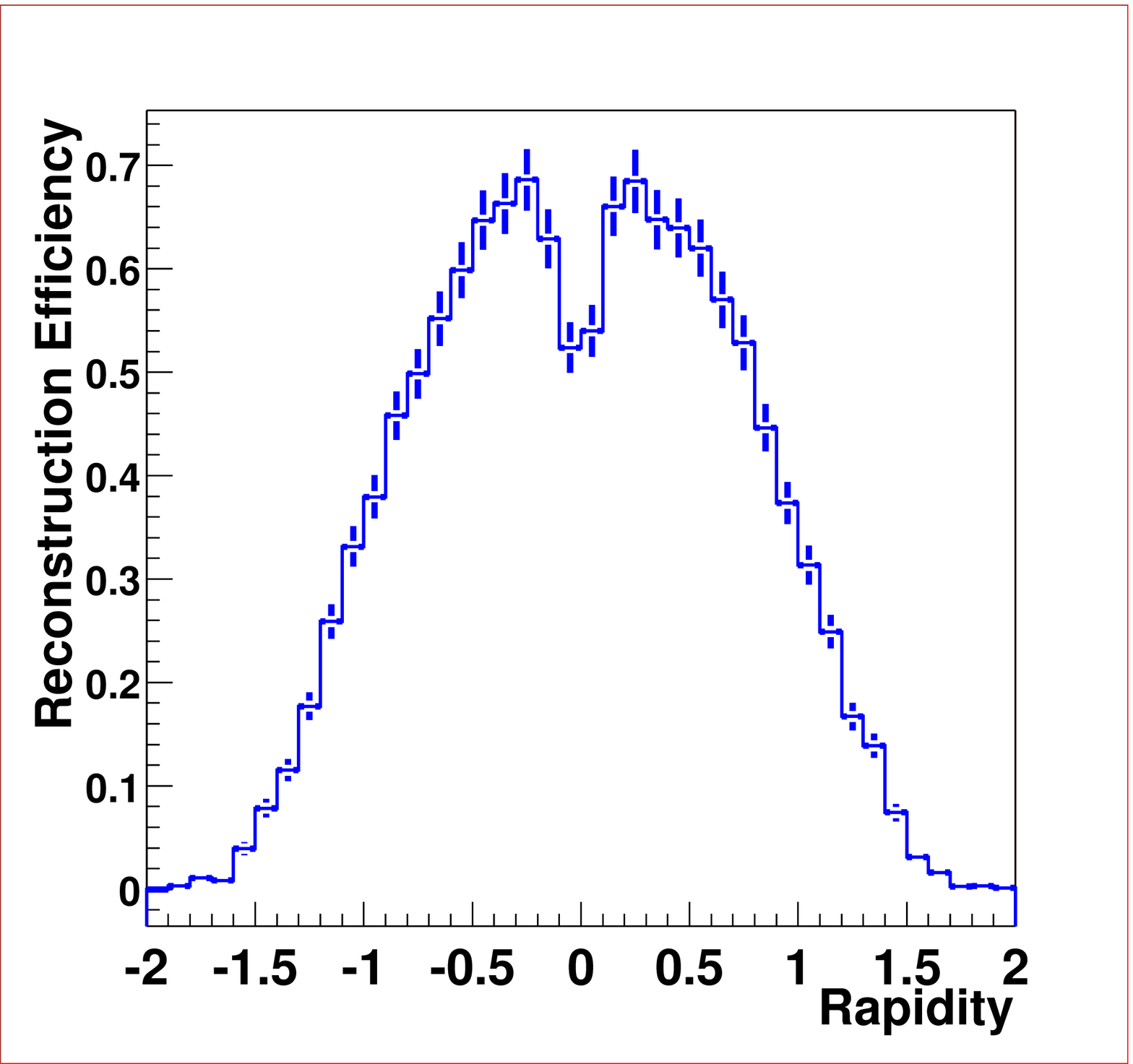}}
\caption{Reconstruction efficiency, including acceptance,
as a function of the rapidity, for the inclusive $\rho^0$ production.}
\label{fig:yEff}
\end{minipage}
\hspace*{0.05\textwidth}
\begin{minipage}{0.45\textwidth}
{\includegraphics[height=.4\textheight,clip=true]{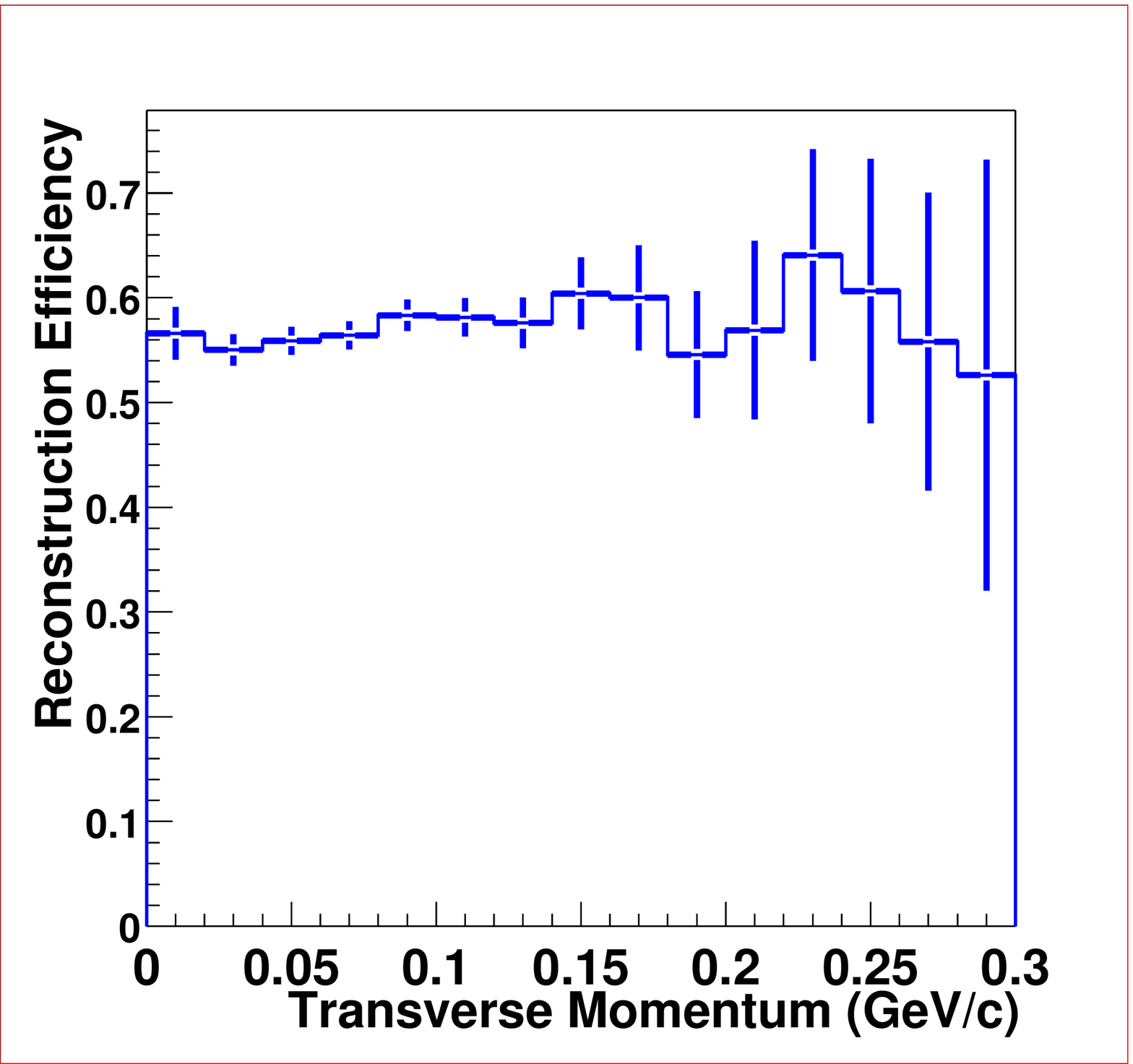}}
\caption{$\rho^0$ meson reconstruction efficiency, including acceptance,
as a function of transverse momentum, before the cut in the opening angle
is applied. Only $\rho^0$ mesons with $|\eta|<$1 were considered.}
\label{fig:ptEff}
\end{minipage}
\end{figure}

\vspace*{-0.5cm}
\noindent

\subsection{Minimum Bias Sample}
The former includes 800k events triggered by requiring simultaneous
signal in both ZDC's.
The analysis of the 'minimum bias' sample, on which
this paper focuses, is optimized to select the following 
reaction: $Au Au \!\rightarrow\! Au^\star Au^\star \rho^0$.
Events are required to have two tracks of opposite charge coming from a 
common vertex.  In addition the vertex should be within 15cm from the beam axis, and
contained in the TPC in the longitudinal direction. Moreover the opening
angle in the transverse plane is required to be larger than 2.7 radians.

A sample of 100K Monte Carlo $AuAu \rightarrow Au Au \rho^0$ events,
generated with STARLIGHT\cite{starlight}, and run through the STAR
simulation and reconstruction chain was utilized to estimate
the reconstruction efficiency for $\rho^0$ inclusive production.
The efficiency, including geometrical acceptance, as a function of the 
$\rho^0$ rapidity is depicted in
Fig.~\ref{fig:yEff}. As can be seen the efficiency peaks at 70\% at
mid rapidity. The dip at rapidity zero is due to the poor vertex
resolution for those events. This is due to the fact that it
is not possible to determine
a common vertex for two tracks perfectly back-to-back.
Fig.~\ref{fig:ptEff} depicts the efficiency as a function of
transverse momentum, before the cut on the angle in the transverse plane
is applied. That cut affects the high end of the distribution, but has
little effect in the overall acceptance, since $\rho^0$ $p_T$ distribution
peaks below 100 MeV/c.

\begin{figure}[!ht]
\includegraphics[height=.5\textheight]{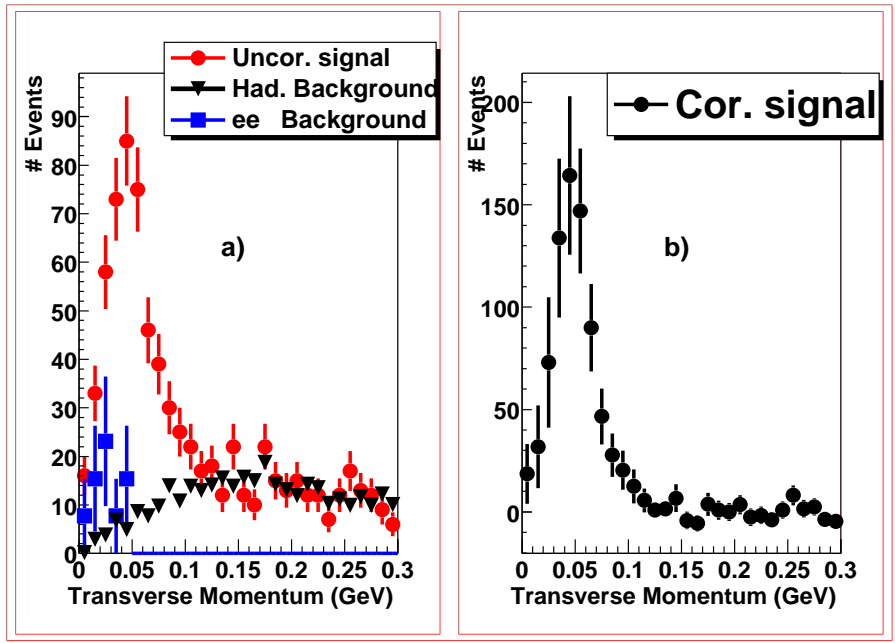}
\caption{a) The uncorrected $p_T$ spectrum for $\rho^0$ candidates
in the minimum bias sample, along with estimates for the hadronic and di-electron
backgrounds.  
b) The $p_T$ spectrum for $\rho^0$ candidates after background subtraction
and correcting for acceptance, reconstruction efficiency, and smearing effects.}
\label{fig:ptCor}
\end{figure}

Fig.~\ref{fig:ptCor}.a shows the uncorrected transverse 
momentum spectrum for the selected events, along with estimated
background from hadronic processes and dielectron events.
A clear peak, the signature for coherent coupling,  can be observed at 
$p_T\!<\!100$~MeV/c. Those events are compatible with coherently produced $\rho^0$ candidates.
The hadronic background is estimated from like-sign combination pairs, 
which is normalized to the signal for  $ p_T \!>\!$ 200 MeV/c. As can be noted, the
hadronic background does not have a peak at low $p_T$. Di-electron background is
estimated by selecting electrons in the kinematic zone where the energy loss dE/dx 
in the TPC provides good separation between electron and pion as explained below.
Fig.~\ref{fig:ptCor}.b shows the corrected transverse momentum after
background subtraction, and correction for 
acceptance, detection efficiency, and smearing effects. 
\begin{figure}[!ht]
\includegraphics[height=.5\textheight]{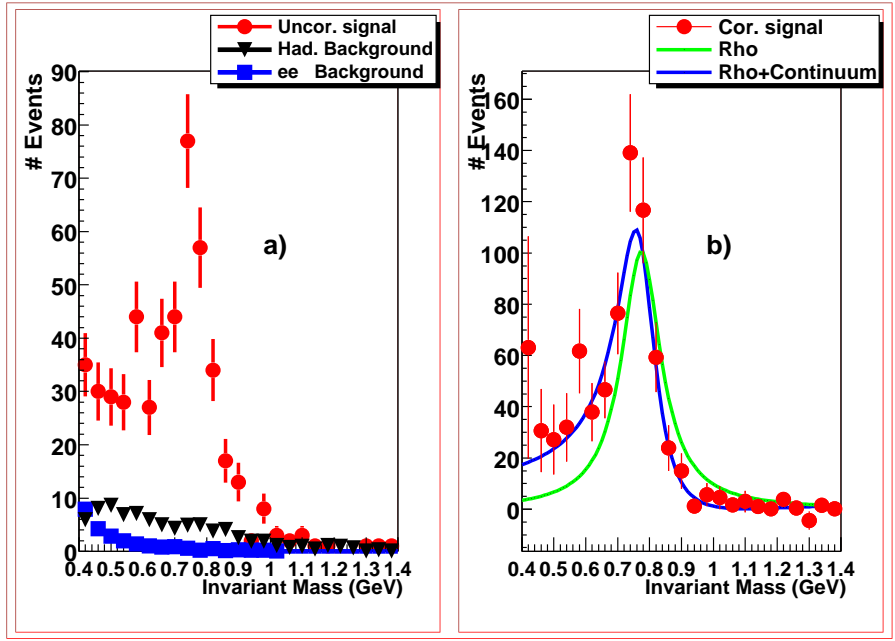}
\caption{a) The uncorrected $m_{\pi\pi}$ spectrum of $\rho^0$ 
candidates with $p_T < 150$ MeV/c, along with estimates 
for hadronic and di-electron
backgrounds. b) The corrected $m_{\pi\pi}$ spectrum of $\rho^0$ candidates with
$p_T < 150$ MeV/c along with the results from a fit to 
equatiion(\ref{eq:rhosigma}).}
\label{fig:massCor}
\end{figure}

Fig.~\ref{fig:massCor}.a depicts the $\pi\pi$ invariant mass distribution for
the events with a total $p_T$ smaller than 150 MeV/c, along with estimates
for hadronic and di-electron backgrounds. As can be observed, 
a clear peak is shown at the $\rho^0$ mass, while backgrounds are concentrated
at low masses. 
The spectrum corrected for background, acceptance, reconstruction efficiency
and smearing is shown in Fig.~\ref{fig:massCor}.b.
The corrected spectrum is well
fit by Eq. \ref{eq:rhosigma}, with $|B/A|$ similar to that found by the ZEUS
collaboration for $\gamma p\rightarrow \rho^0
p$\cite{ZEUS}.  

\subsection{Topology Sample}
%%% ====  Topology trigger 
For the analysis of the process $AuAu \rightarrow AuAu \rho^0$ 
a data sample selected with a low-multiplicity topology trigger
was utilized.
In the level 0 trigger, the central trigger barrel was divided into
16 coarse pixels.
These pixels cover half a unit in rapidity, and $90^0$ in 
the transverse plane.
Hits in opposite pixels were required, while pixels in the top and the bottom 
acted as a veto to supress cosmic rays.  No requirement on the ZDC signals
was imposed. A high level trigger\cite{L3}
further removed background. The STAR collaboration collected about 30k 
events in 7 hours, with this configuration. The $\rho^0$ candidates from this 
data set have a transverse momentum and an 
invariant mass distributions similar to the ones already shown in 
Figures ~\ref{fig:ptCor} and \ref{fig:massCor}. Therefore, they are also 
characterized by a $p_T$ distribution peaked below 100 MeV/c, and  
and events, around 300, clustered around the $\rho^0$ mass. 
In contrast to the minimum bias data, the topology triggered data had almost
no energy deposition in the ZDC consistent with the two gold nuclei remaining
in their ground state.

\subsection{Electron-Positron pair production}
\label{di-electron}
\begin{figure}[!ht]
\vspace*{-0.3cm}
\resizebox{.45\textwidth}{!}
{\includegraphics[height=.5\textheight,angle=270]{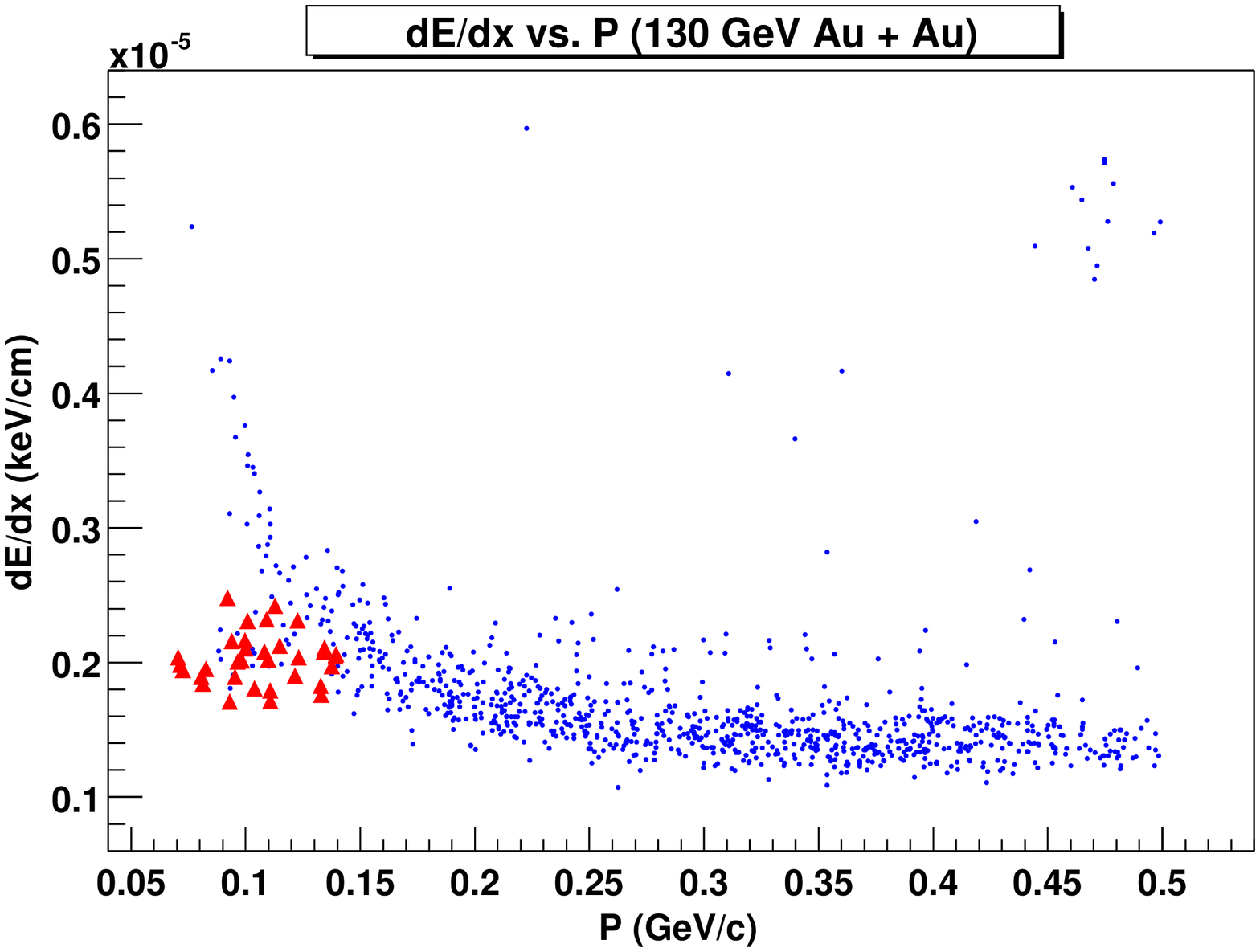}}
\resizebox{.45\textwidth}{!}
{\includegraphics[height=.5\textheight,angle=270]{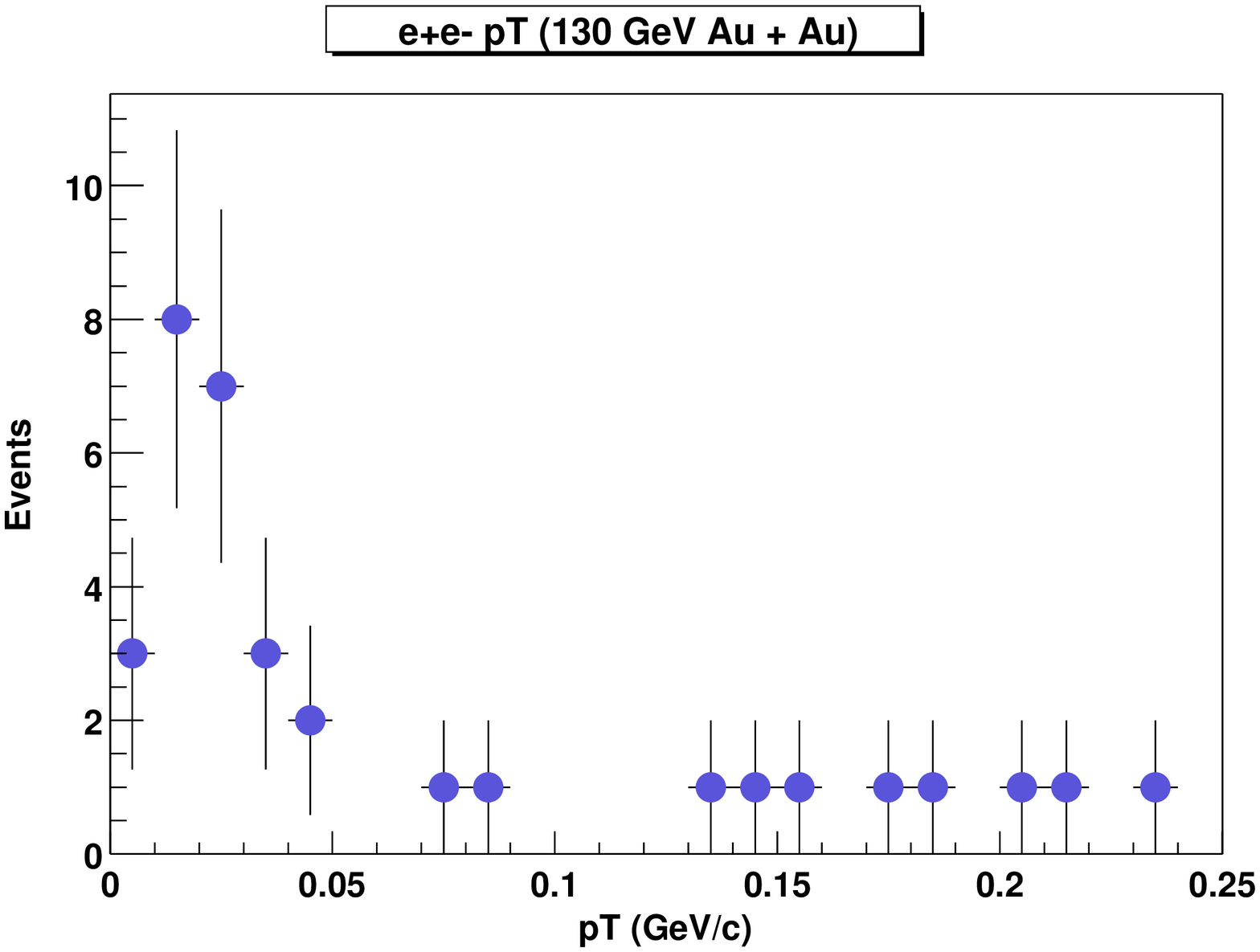}}
\caption{(a) Energy loss $dE/dx$ of tracks in the 2-track, minimum bias data;
triangles indicate events where both particles are identified as electrons. 
(b) The $p_T$ spectrum for  identified $e^+e^- pairs$. \label{fig:electrons} }
\vspace*{-0.5cm}
\end{figure}

%%%% Di-electron
Two-photon interactions include the purely electromagnetic process of electron-positron 
pair production as well as  single and multiple meson production.  The coupling $Z\alpha$
($0.6$ for $Au$) is large, hence  $e^+e^-$ pair production is an important
probe of quantum electrodynamics in strong fields~\cite{baurrev}.
At momenta below $140$ MeV/c, $e^+e^-$ pairs are identified by  
their energy loss in the TPC as shown for the minimum bias data sample
in Fig.~\ref{fig:electrons}a. Fig.~\ref{fig:electrons}b shows the $p_T$
spectrum for  identified $e^+e^-$ pairs;  a clear peak at $p_T \!< \!50$ MeV/c
identifies the process $AuAu \rightarrow Au^\star Au^\star e^+e^-$. 

%% Summary
\section{Summary}
In summary,  for the first time, 
exclusive $\rho^0$ production  $AuAu \!\rightarrow\! Au Au \rho^0$ and  $\rho^0$
production with and without nuclear breakup was 
observed in ultra-peripheral heavy  ion collisions. 
The $\rho^0$ are produced at small perpendicular momentum,
showing their coherent coupling to both nuclei. 
In addition, the coherent electromagnetic process 
$Au Au \rightarrow Au^\star Au^\star e^+ e^-$ was observed.

\end{document}